\documentclass[a4paper,aps,pre,twocolumn,groupedaddress,showkeys,showpacs,floats,floatats,floatfix]{revtex4-1}
\usepackage[utf8]{inputenc}
\usepackage[english]{babel}
\usepackage{graphicx}
\usepackage{dcolumn} 
\usepackage{bm}
\usepackage{natbib}
\usepackage{latexsym}
\usepackage{mathrsfs}
\usepackage{amssymb}
\usepackage{amsmath}
\usepackage{amscd}
\usepackage{color}
\usepackage{pifont}
\usepackage{pstricks,pst-node,pst-text,pst-3d}
\usepackage{verbatim}
\usepackage{ulem}
\usepackage[T1]{fontenc}
\usepackage{hyperref}
\definecolor{blue-violet}{rgb}{0.54, 0.17, 0.89}

\begin{document}

\title{Decay of distance autocorrelation and Lyapunov exponents}
\author{C.~F.~O.~Mendes$^1$, R.~M.~da Silva$^1$ and M.~W.~Beims$^{1,2}$}
%

\affiliation{$^1$Departamento de F\'\i sica, Universidade Federal do Paran\'a,
         81531-980 Curitiba, PR, Brazil}
\affiliation{$^2$Max-Planck Institute for the Physics of Complex Systems,
          N\"othnitzer Strasse 38, 01187, Dresden, Germany}

\date{\today}
%
\begin{abstract}
This work presents numerical evidences that for discrete dynamical systems with 
one positive Lyapunov exponent the decay of the distance autocorrelation is always 
related to the Lyapunov exponent. Distinct decay laws for the distance autocorrelation 
are observed for different 
systems, namely exponential decays for the quadratic map, logarithmic for the Hénon map 
and power-law for the conservative standard map. In all these cases the decay 
exponent is close to the positive Lyapunov exponent. For hyperbolic conservative 
systems the power-law decay of the distance autocorrelation tends to be 
guided by the smallest Lyapunov exponent.
\end{abstract}
%
\maketitle

\section{Introduction}
\label{intro}

The understanding of decay of correlations in dynamical systems is of crucial 
relevance in nonequilibrium statistical physics. It may furnish the qualitative and 
quantitative knowledge about, for example, relaxation processes in complex
systems \cite{vulpiani90}. In such complex systems a chaotic behaviour is common and a 
relation between the decay of correlations with positive Lyapunov exponents (LEs) is 
intuitively expected. From the theoretical point of view such relation was studied before 
using the standard autocorrelation function between two observables, as for example, 
in 1D  dissipative systems \cite{politi88,just13-1,just13-2}, in area preserving maps 
\cite{cary83}, in 2D maps \cite{politi88} with random perturbations \cite{blumen18}, and 
conjectured that for higher-dimensional systems the smallest positive LE
is an upper bound for the  decay of correlation \cite{collet04}.

In this work we study the decay {of a more recently  \cite{Szek07}  proposed 
correlation}, named  the {\it distance correlation} or {\it covariance correlation}. It was
proposed for testing joint independence of random vectors in arbitrary dimensions
{and} has received attention in applied statistics \cite{Szek09,
Szek12,Szek13,Szek14,Szek17}, in quantum-classical transitions analysis for ratchet systems 
\cite{Beims}, in genetic risk problems \cite{Kong}, in functional brain connectivity \cite{geerlig} 
and for the parameter identification of nonlinear systems \cite{Solar}. It was shown 
recently \cite{CFMWB} that the distance correlation is able to detect noise induced escape 
times from regular and chaotic attractors, to describe mixing properties between chaotic 
trajectories and to detect properties related to the linear stability of orbital points. 
{However, no direct relation to the LE was achieved.}  

The crucial issue is to extract 
from the distance correlation relevant  properties related to nonlinear dynamics hidden 
in time series. {Inspired by this, the present work aims to give a qualitative and 
quantitative connection between the distance correlation and LEs.}
Being more specific, we show that the positive LE is {closely related to}
the decay of the distance correlation calculated between the time series and 
itself delayed in time, {named here as {\it distance autocorrelation} $(DA)$}. 
{Results are presented for two dissipative maps, the quadratic and the H\'enon 
maps and for the conservative standard map and coupled standard maps. Distinct 
qualitative decays of $DA$ as a function of the delay time $\Delta t$ are observed for these 
maps, ranging from exponential, logarithm to power-law. In all cases the decay exponents are 
close to the LEs. Exceptions only occur for hyperchaotic systems, for which the decay 
exponents tend to follow the smallest LEs.}

The paper is organized as follows. While in Sec.~\ref{DCm} we review the general definition of 
the distance correlation, Sec.~\ref{delay} describes the $DA$ used in this work.
Section \ref{period} shows how the $DA$ can be used to  determine the periodicity of a time 
series and Sec.~\ref{LE} shows its relation to the LEs from the quadratic map. In Sec.~\ref{high} 
we test our findings for 
higher-dimensional systems, namely the H\'enon map and (coupled) standard maps. Section 
\ref{conclusions} summarizes our main results.

\section{Distance correlation}
\label{DCm}

In this Section we summary the main properties of the distance correlation \cite{Szek07} as a 
statistical measure of dependence between random vectors which is based on Euclidean distances. 
We present here only the computational procedure to determine the distance correlation. 
For more details about this procedure and original definitions we refer 
the readers to \cite{Szek07}.

Assume two random samples $(\boldsymbol{X},\boldsymbol{Y}) = \{(X_{k},Y_{k}):k=1,\dots,N\}$ 
for $N \geq 2$ and $X \in \mathbb{R}^{s}, Y \in \mathbb{R}^{t}$ with $s$ and $t$ integers. The 
{\it empirical distance correlation} is given by the expression 
\begin{equation}
\label{dc}
DC_{N}(\boldsymbol{X},\boldsymbol{Y})=\frac{\sigma_{N}(\boldsymbol{X},\boldsymbol{Y})}{
\sqrt{\sigma_{N}(\boldsymbol{X})\sigma_{N}(\boldsymbol{Y})}},
\end{equation}
where the {\it empirical distance covariance} $\sigma_{N}(\boldsymbol{X},\boldsymbol{Y})$
for a joint random sample $(\boldsymbol{X}, \boldsymbol{Y})$ is defined by the expression
\begin{equation}
\label{xy} 
\sigma_{N}(\boldsymbol{X},\boldsymbol{Y}) = \frac{1}{N}\left( \sum_{i,j=1}^{N} A_{ij}B_{ij}
\right)^{1/2},
\end{equation}
where $A$ and $B$ are matrices. The {\it empirical distance variance} for a random 
sample $\boldsymbol{X}$ is given by
\begin{equation}
\label{} 
\sigma_{N}(\boldsymbol{X}) = \frac{1}{N}\left( \sum_{i,j=1}^{N} A^{2}_{ij}\right)^{1/2},
\end{equation}
and for a random sample $\boldsymbol{Y}$
\begin{equation}
\label{} 
\sigma_{N}(\boldsymbol{Y}) = \frac{1}{N}\left(\sum_{i,j=1}^{N} B^{2}_{ij}\right)^{1/2}.
\end{equation}
For $X \in \mathbb{R}^{s}$ with $i=1, \ldots,N$ and $j=1, \ldots, N$, the matrix $A$ is 
obtained from 
\begin{equation}
\label{}
A_{ij} = a_{ij} - \bar{a}_{i.} - \bar{a}_{.j} + \bar{a}_{..},
\end{equation}
where $a_{ij} = |X_{i} - X_{j}|$ is the Euclidean norm of the distance between the elements 
of the sample, $\bar{a}_{i.} = \frac{1}{N} \sum_{j=1}^{N} a_{ij}$ and $\bar{a}_{.j} = \frac{1}{N} 
\sum_{i=1}^{N} a_{ij}$ are the arithmetic mean of the rows and columns, respectively, and 
 $\bar{a}_{..} = \frac{1}{N^{2}} \sum_{i,j=1}^{N}  a_{ij}$ is the general mean. Similarly to $Y \in 
\mathbb{R}^{t}$ defined by $i=1, \ldots, N$ and $j=1, \ldots, N$, we also define the matrix
\begin{equation}
\label{ma}
B_{ij} = b_{ij} - \bar{b}_{i.} - \bar{b}_{.j} + \bar{b}_{..},
\end{equation}
where the terms $b_{ij}$, $\bar{b}_{i.}$, $\bar{b}_{.j}$ and $\bar{b}_{..}$ are similar to 
those for the matrix $A_{ij}$. 

The $DC_{N}(\boldsymbol{X},\boldsymbol{Y})$ is defined inside the interval $[0,1]$ and 
its main characteristic is that it will be zero if and only if the random vectors are independent 
\cite{Szek07, Szek09, Szek12, Szek13, Szek14}. In addition, it is easy to check that 
$DC_{N}(\boldsymbol{X},\boldsymbol{Y})$ is scale independent. This means that 
samples $\boldsymbol{X}$ and $\boldsymbol{Y}$ can be multiplied by $\alpha$ and
$\beta$,  with $\alpha,\beta \in \mathbb{R}$, and $DC_{N}(\boldsymbol{X},\boldsymbol{Y})$ 
remains unaltered. The $DC_{N}(\boldsymbol{X},\boldsymbol{Y})$ only requires a series
of data in order to be analyzed. This can be crucial when analyzing experimental data.

\section{Distance autocorrelation}
\label{delay}

Defining matrix $B$ to be equal to matrix A with a delay $\Delta t$, the matrix elements are related by 
$B_{i,j}=A_{i+\Delta t,j+\Delta t}$. {In the present work we always consider time series,
so that $\Delta t$ is a time delay.} Using increasing values of $\Delta t$ this is a discrete 
convolution between $A$ and $A_{i+\Delta t,j+\Delta t}$ and we expect to obtain some relevant 
information about the dynamics of the original time series as a function of $\Delta t$. In 
other words, for the calculation of $DA$, we use here only {\it one} time series of states 
of a particular initial condition (IC) to compose the sample $\{\boldsymbol{X}\}$. From this 
time series, we obtain the data set $\{\boldsymbol{Y}\}$ by only  shifting the states of 
$\{\boldsymbol{X}\}$ on time. Thus, we must consider the following representation for the 
joint sequence $(\boldsymbol{X},\boldsymbol{Y})$
\begin{flalign}	 
\boldsymbol{X} \Rightarrow &(X_{n}:{X_{1},X_{2},X_3,\dots,X_{N}}),&\nonumber\\
\boldsymbol{Y} =\boldsymbol{X'} \Rightarrow &(X_{n}:{X_{1+\Delta t},X_{2+\Delta t},
X_{3+\Delta t},\dots,X_{N+\Delta t}}).&\nonumber
\end{flalign}
Here $n$ is an integer referring the times for which the variables $(\boldsymbol{X,Y})$ 
were determined. For maps, $1 \le \Delta t \le N$ is a positive integer. 
{Thus, distance autocorrelation function is then defined by
\begin{equation}
\label{dc}
DA = DC_{N}(\boldsymbol{X},\boldsymbol{X'})=\frac{\sigma_{N}(\boldsymbol{X},\boldsymbol{X'})}{
\sqrt{\sigma_{N}(\boldsymbol{X})\sigma_{N}(\boldsymbol{X'})}}.
\end{equation}
}

\section{Detecting periodicity}
\label{period}
It is easy to realize that the $DA$ calculated between samples $\boldsymbol{X}$ and 
$\boldsymbol{X'}$ will be $DA=1$ when $\Delta t=m$, with $m$ being the periodicity of the 
time series. To check this obvious property we present numerically results only for one 
time discrete dynamical system, the quadratic map.

\begin{figure}[!h]
\begin {center} 
  \includegraphics [width=8.1cm,height=4.6cm]{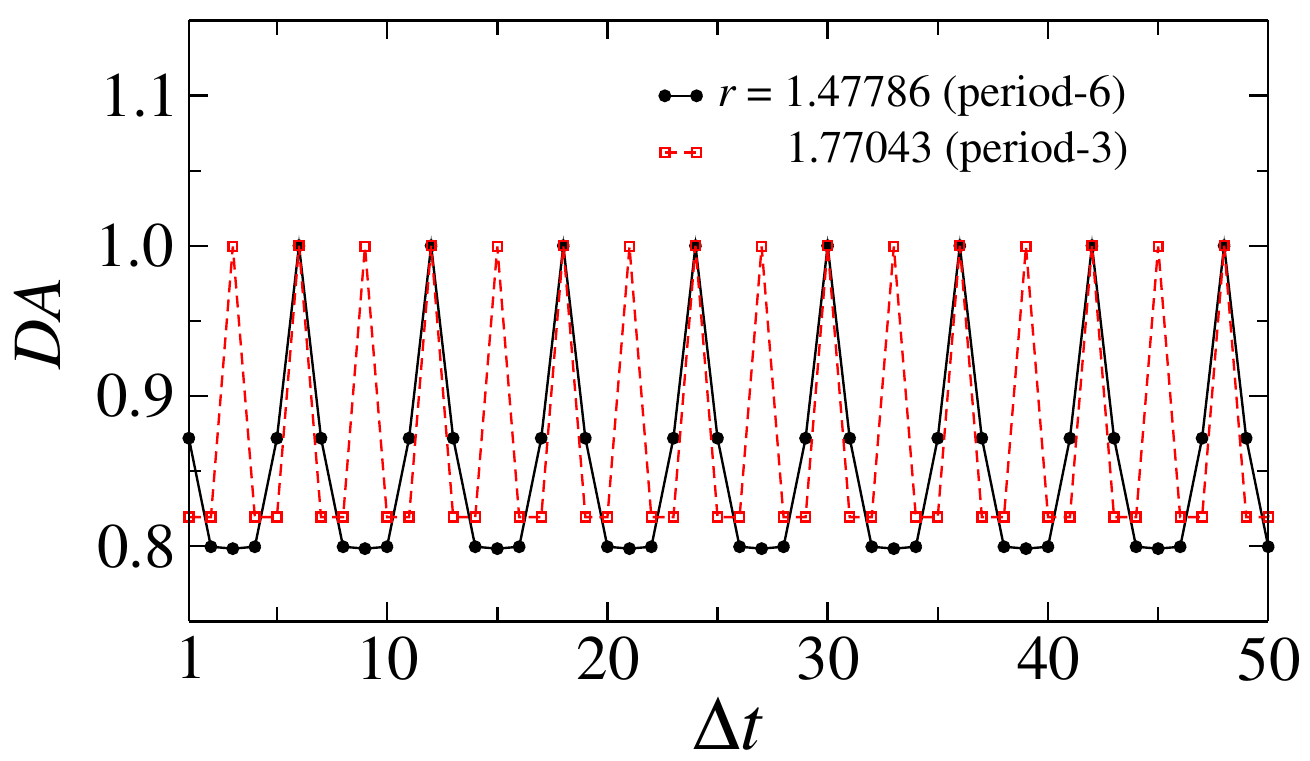}
  \caption{$DA$ as a function of $\Delta t$ for the QM 
  (\ref{map}) considering two periodic windows: 
  $r=1.47786$ (period-$6$) and $r=1.77043$ (period-$3$).}
  \label{fig1}
\end{center}
\end{figure}

{\it Quadratic map} (QM). The quadratic map is a one-dimensional dynamical system given by 
the expression \cite{allig, topol}
\begin{equation}
\label{map} 
x_{n + 1} = r-x_{n}^{2},
\end{equation}
where $r$ is a control parameter that belongs to the range $[-0.25,2]$, $n =
0,1,2,3, \ldots$ the integer which represents the number of iterations and $x_{n} \in 
[-2,2]$ is the state of the system at time $n$. Results for $DA$ as a function of $\Delta t$ 
are presented in Fig.~\ref{fig1}. $DA$ is shown for two periodic windows located at the 
parameters: $r=1.47786$ (period-6) and $r=1.77043$ (period-3). We notice that in both cases 
$DA=1$ when $\Delta t=6$ and $\Delta t=3$, respectively. This means that $DA$ furnishes a 
measure of how much the time series repeats itself. Since this is an evident property  
we will not show it in other dynamical systems.

\section{Relation to Lyapunov exponents}
\label{LE}

In this Section we show that the $DA$ calculated in Sec.~\ref{delay} is 
related to the maximal LE. We mention that while the LE gives the exponential 
divergence of nearby trajectories in the linear approximation, the $DA$ may contain in itself 
nonlinear extensions of this property and is not restricted to furnishes solely the usual LE.

\begin{figure}[!t]
\begin{center} 
  \includegraphics[width=8.6cm,height=4.3cm]{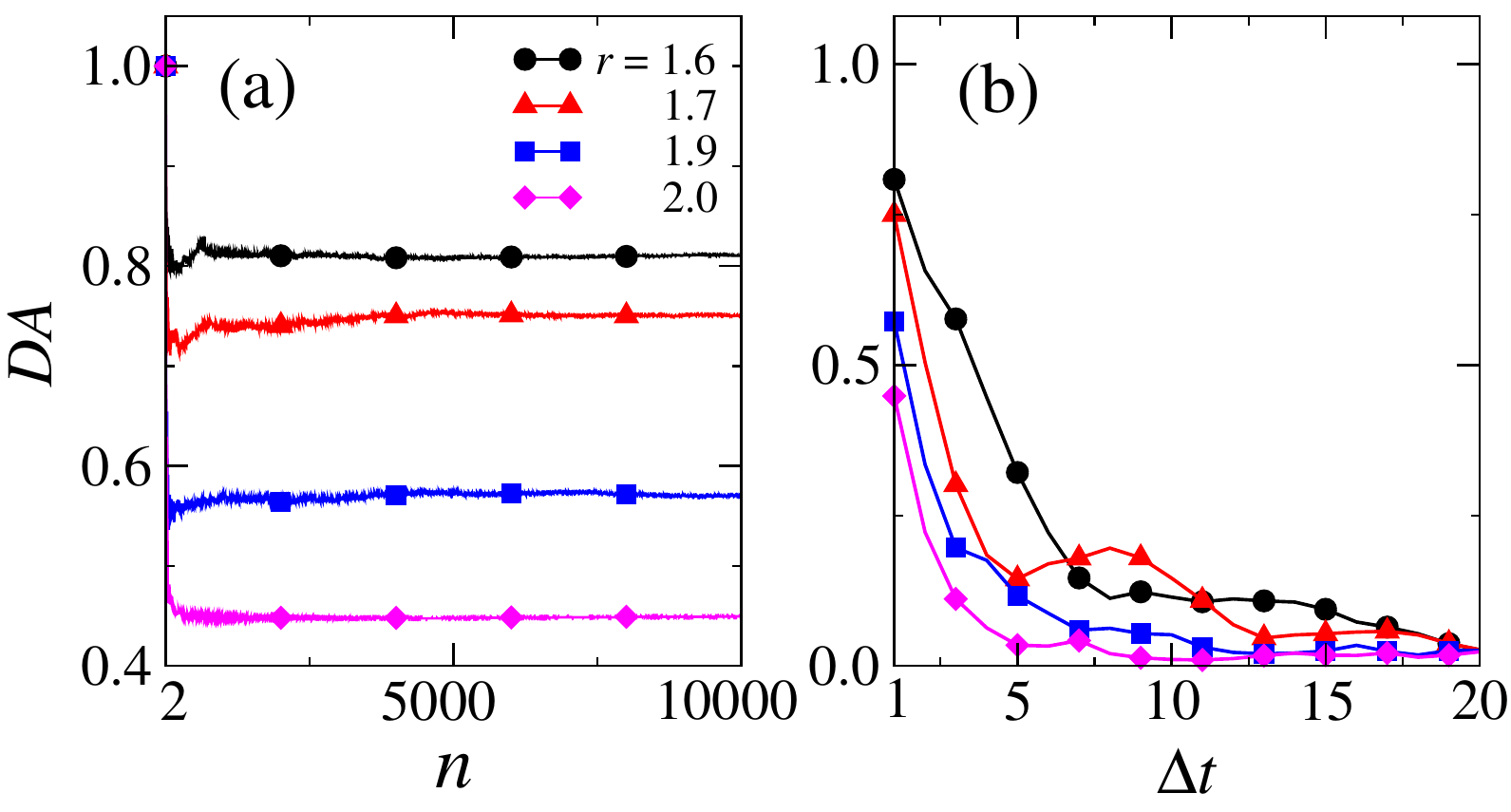}
  \caption{(a) Time evolution of $DA$ for four parameters corresponding to the chaotic regime 
  of the QM (\ref{map}). The parameters are $r=1.6$, $1.7$, $1.9$ and $2.0$. The IC used is 
  $x_{0}=0.1$, and the displacement of the states between the samples is $\Delta t=1$. (b) 
  $DA$ as a function of $\Delta t$ for the same parameters from (a).} 
  \label{fig2}
\end{center}
\end{figure}

To start the discussion we have firstly to guarantee that the $DA$ converges to a reasonable
value for a given $\Delta t$. This is presented in Fig.\ref{fig2}(a) for the QM with $r=1.6$, 
$1.7$, $1.9$ and $2.0$. In this case, $\Delta t=1$. This shows that for each value of $r$ the 
$DA$ converges asymptotically, as a function of $n$, to distinct finite values. In other 
words, even for a chaotic trajectory, $DA$ does not converges to zero. Such situation 
changes drastically when $DC_{N}(\boldsymbol{X},\boldsymbol{Y})$ is calculated between 
two distinct ICs inside the chaotic regime.  This leads to mixing properties for which
$DC_{N}(\boldsymbol{X},\boldsymbol{Y})$  tends to zero with $n^{-1/2}$, as shown in 
\cite{CFMWB}.

The next step is to show how such asymptotic values of $DA$ change with $\Delta t$. This 
is presented in Fig.~\ref{fig2}(b), which displays $DA$ as a function of $\Delta t$ for 
parameters inside  the chaotic regimes of the QM, namely $r=1.6$, $1.7$, $1.9$ and $2.0$. No 
values $DA=1$ are observed since the chaotic trajectory is not periodic anymore. We note that 
all $DA$ curves  converge to small, but finite values when $\Delta t=20$.

\begin{figure}[!t]
\begin{center} 
  \includegraphics [width=8cm,height=7.5cm]{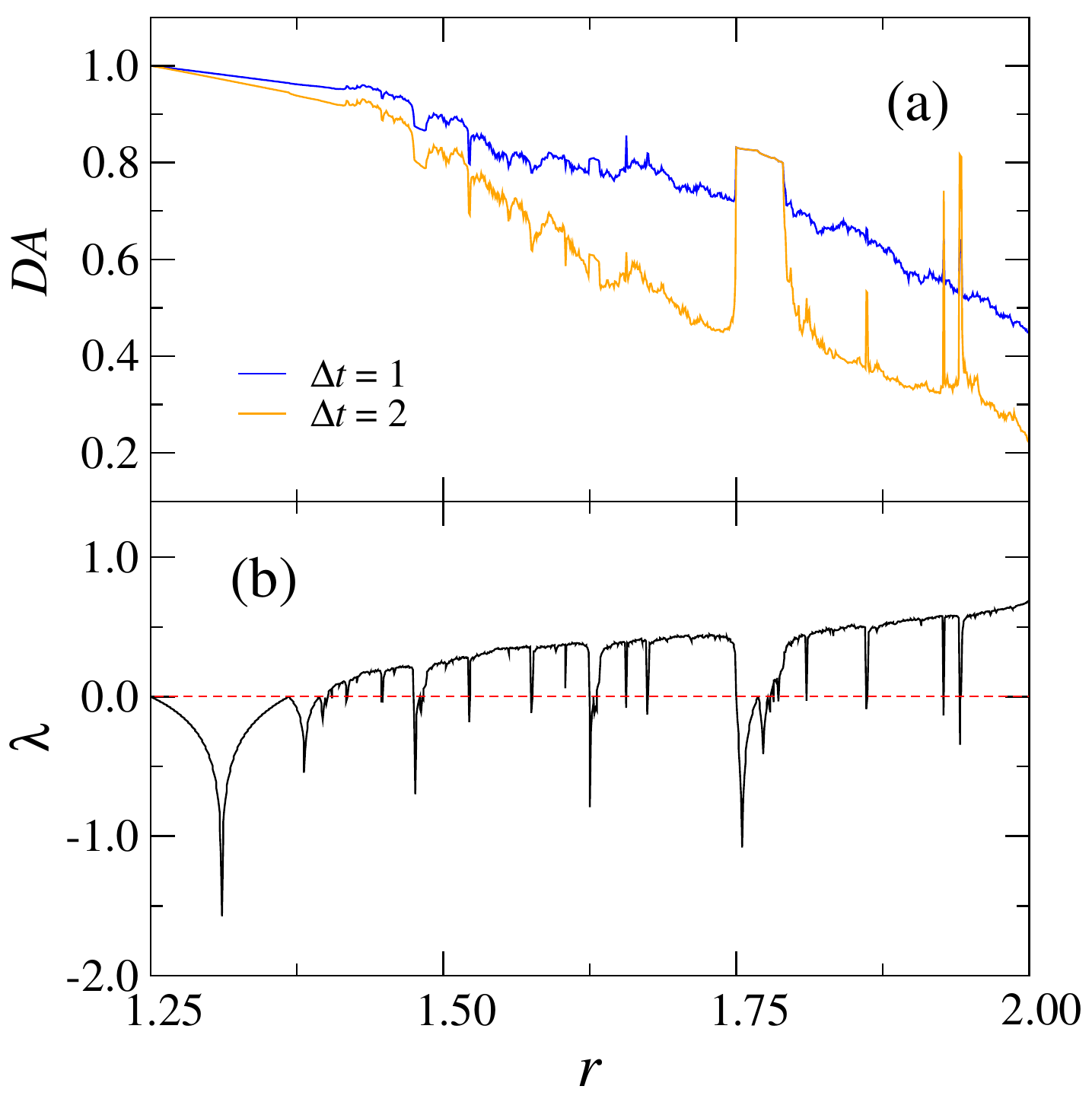}
  \caption{(a) $DA$ for $\Delta t=1$ (blue) and $\Delta t=2$ (orange) and (b) LE, both as 
  a function of $r$. The initial condition used is $x_{0}=0.1$ and the parameter $r$ is 
  divided into $10^3$ equally spaced parts.} 
  \label{fig3}
\end{center}
\end{figure}

From Fig.~\ref{fig2}(a) we observe that when increasing $r$
 the corresponding $DA$ decrease. Even though we expect that for time series 
with larger LEs the $DA$ is smaller, we wonder if there is an analytical relation between both. 
To find such relation we firstly determine $DA$ for many values of parameters inside the interval 
$1.25 \le r \le 2.0$ of the QM. This is shown in Fig. \ref{fig3}(a). In this case we considered just 
two time delays, $\Delta t=1$ (blue curve) and $\Delta t=2$ (orange curve) from a total iteration 
time of $n=10^{4}$. A direct comparison between $DA$ and the LE ($\lambda$) from the QM, shown in 
Fig.\ref{fig3}(b), allows us to realize that, besides the regions with periodic motion, $DA$ 
decreases while $\lambda$ increases with the increment of $r$. It becomes clear that 
$DA$ is able to identify the periodic windows that occur at specific values of $r$.
%
\begin{figure}[!b]
\begin{center} 
  \includegraphics[width=7.2cm,height=6.2cm]{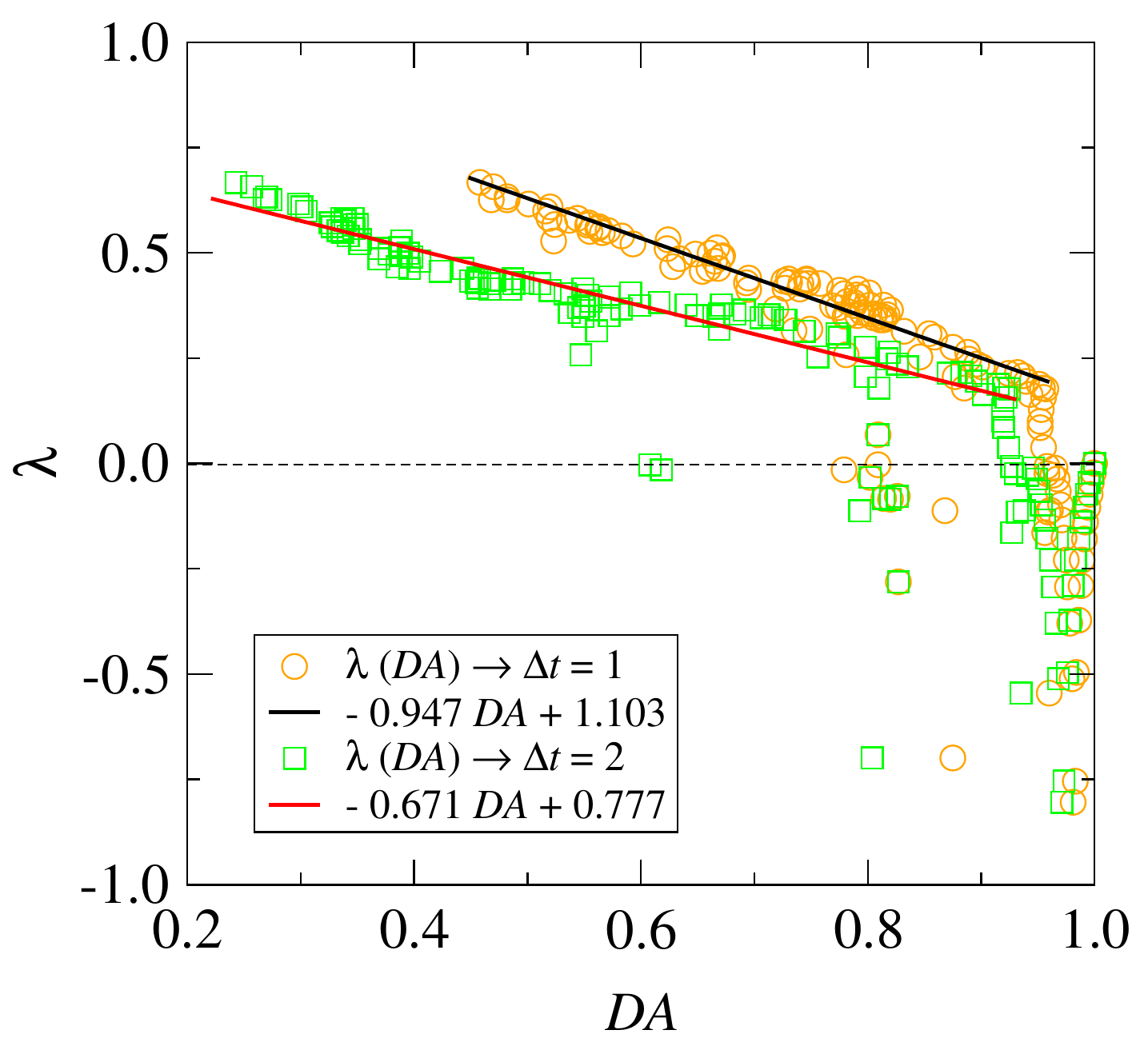}
  \caption{LE as a function of $DA$ for the cases $\Delta t=1$ and $\Delta t=2$. The functions 
  for the adjustments are given in the inset.} 
  \label{fig4}
\end{center}
\end{figure}

{Now we combine data for $DA$ from Fig.~\ref{fig3}(a) with those for
$\lambda$ from Fig.~\ref{fig3}(b). This is plotted in
Fig.~\ref{fig4}, and shows} the relation between the LE and $DA$ for $\Delta t=1,2$. 
Clearly almost all positive LEs belong to decreasing lines when increasing $DA$. Linear
adjusted lines are given in the inset of this figure. Since the above analysis depends on the 
specific $\Delta t$ used, it is not appropriate to extract general nonlinear properties from 
$DA$. Thus, we move forward and argue that there should exist some relation between the LEs 
and the time decay of $DA$ as a function of $\Delta t$. This argument is reasonable due to the 
fact that the $DA$ calculated for the time series with itself for distinct values of $\Delta t$ 
is related to the convolution between both series.
\begin{figure}[!b]
\begin{center} 
  \includegraphics[width=6.8cm,height=9.2cm]{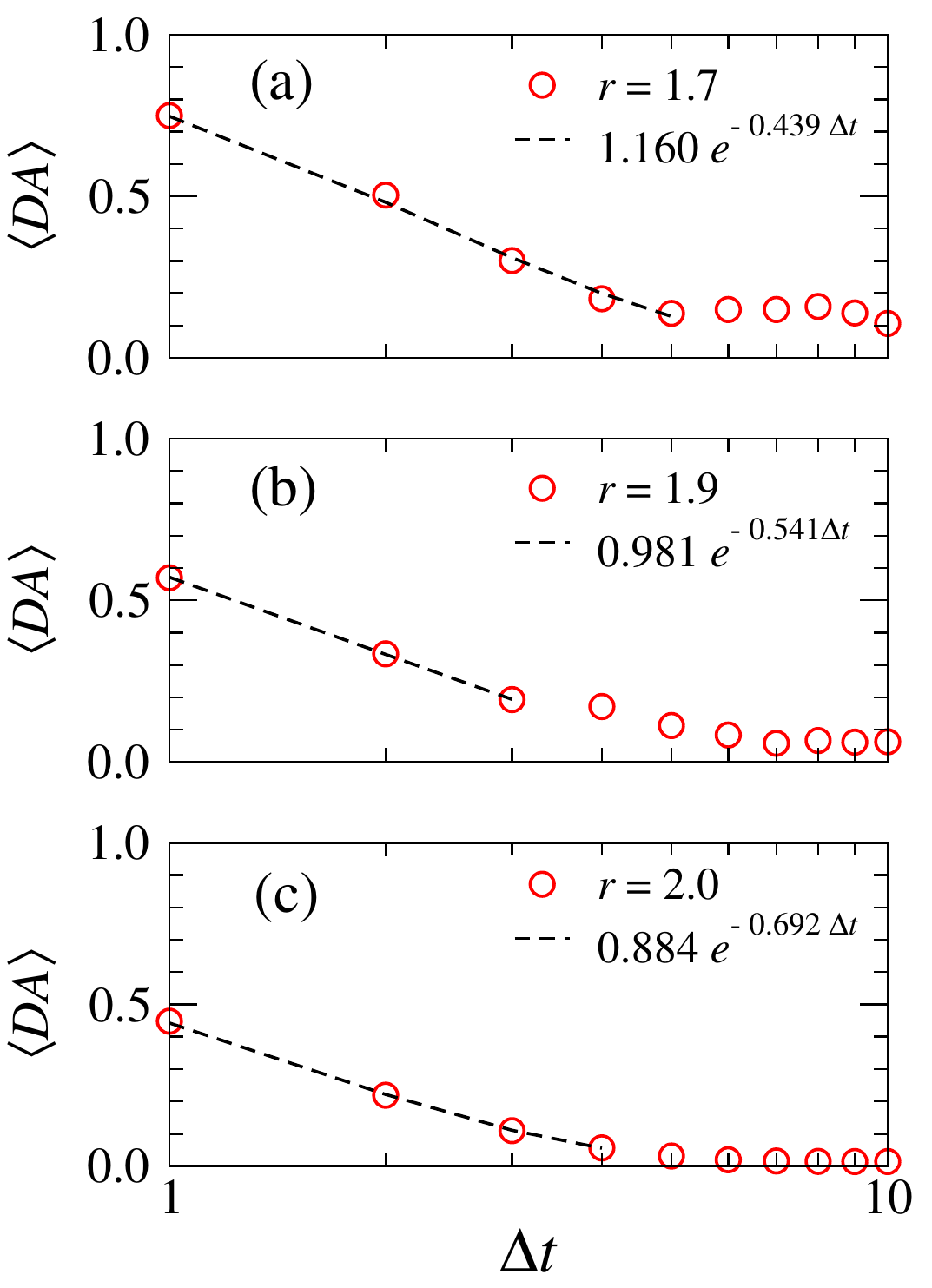}
  \caption{Curves plotted in semi-log scale of the $\langle DA 
  \rangle$ as a function of $\Delta t$ for (a) $r=1.7$, (b) $r=1.9$ and (c) $r=2.0$. 
  The functions for the adjustments (dashed black lines) are given.} 
  \label{fig5}
\end{center}
\end{figure}

\begin{figure}[!t]
\begin{center} 
  \includegraphics [width=7.5cm,height=8.0cm]{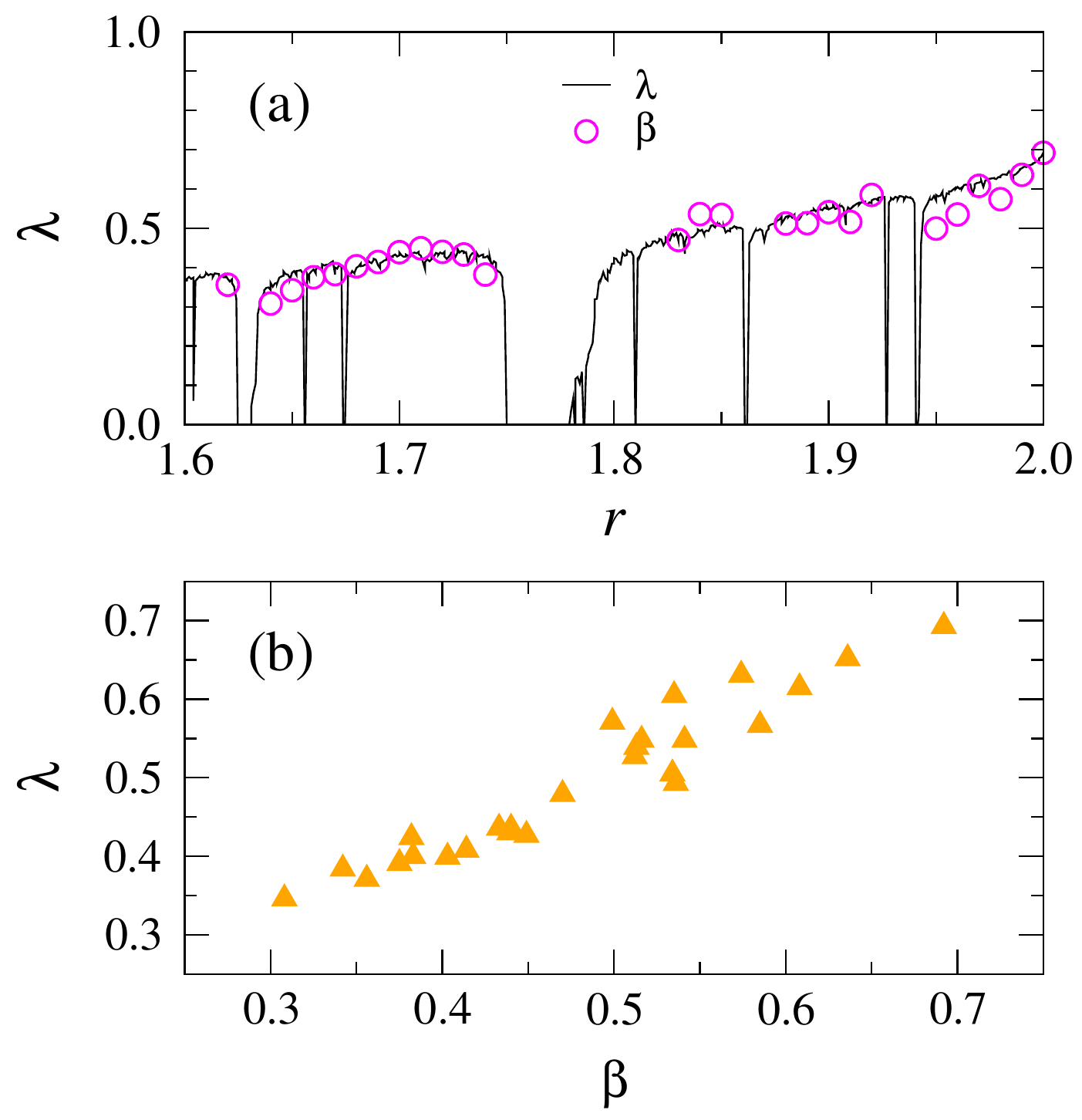}
  \caption{(a) LE and the adjustment parameter $\beta$ in function of $r$ and (b) 
  relationship between $\lambda$ and $\beta$.}
\label{fig6}
\end{center}
\end{figure}

To check this, we obtained the samples $\{\boldsymbol{X}=x_{n+1}:x_{1},x_{2},\dots,x_{N}\}$ 
and $\{\boldsymbol{X'}=x_{n+1}:x_{1+\Delta t},x_{2+\Delta t},\dots,x_{N+\Delta t}\}$ by 
iterating the QM (\ref{map}) $n=10^4$ times after to discard the first $5 \times 10^{3}$ 
transitory iterations. Thus, to determine $DA$ we calculate an average over  $5\times 10^{3}$ 
ICs randomly chosen with equal probability inside the interval $[-2,2]$  to reach a satisfactory 
level of accuracy in the asymptotic $DA$ values. We write the mean distance autocorrelation as 
$\langle DA \rangle$. Figure \ref{fig5} shows $\langle DA \rangle$ (red circles) as function 
of $\Delta t$ in semi-log scale for three different values of the parameter $r$ in the chaotic 
regime. We consider the displacement only in the interval $1 \le \Delta t \le 10$. In 
Fig.~\ref{fig5}(a) we display the case $r=1.7$, which has a LE $\lambda=0.438$.
In Figs.~\ref{fig5}(b) and \ref{fig5}(c) the cases $r=1.9$ and $r=2.0$ are shown, for which 
$\lambda=0.548$ and $\lambda=0.693$ are found, 
respectively. Straight lines in these curves can be fitted by exponential decays of the type 
$\langle DA\rangle \propto \alpha \exp{(-\beta \Delta t)}$, where $\alpha$ and $\beta$ are 
adjustment parameters. Remarkably, the parameters $\beta=0.439$, $\beta=0.541$ and 
$\beta=0.692$ 
provides similar values as the corresponding LEs $\lambda=0.438$, $\lambda=0.548$ and 
$\lambda=0.693$, as can be observed from the adjusted curves (dashed black line) 
in Fig.~\ref{fig5}. We have checked the relation $\beta\approx\lambda$ for other parameters 
of the QM. This is displayed in Fig.~\ref{fig6}(a), which shows the LE 
$\lambda$ (continuous black line) and $\beta$ (circles) as a function of $r$.  From this 
figure, we can conclude that the relation $\beta\approx\lambda$ is valid for the values 
of $r$ that lead to chaotic dynamics. Figure \ref{fig6}(b) {combines data for $\lambda$ and
$\beta$ in the plane}  $\lambda\times\beta$. These results show that the decay of 
$DA$ as a function of $\Delta t$ is closely related to the LE of the dynamical system. We have 
also tried to increase the number of samples, {\it i.~e.~} iteration $n$ and ICs, but no 
relevant changes were observed.

\section{Extension to higher-dimensional systems}
\label{high}

In the following we analyze the relation of the decay of $DA$ with the LE in 
nonlinear systems {with more dimensions}. We start with the dissipative 
H\'enon map and then consider the conservative (coupled) standard map.

\begin{figure}[!b]
\begin{center} 
\includegraphics[width=8.72cm,height=9.5cm]{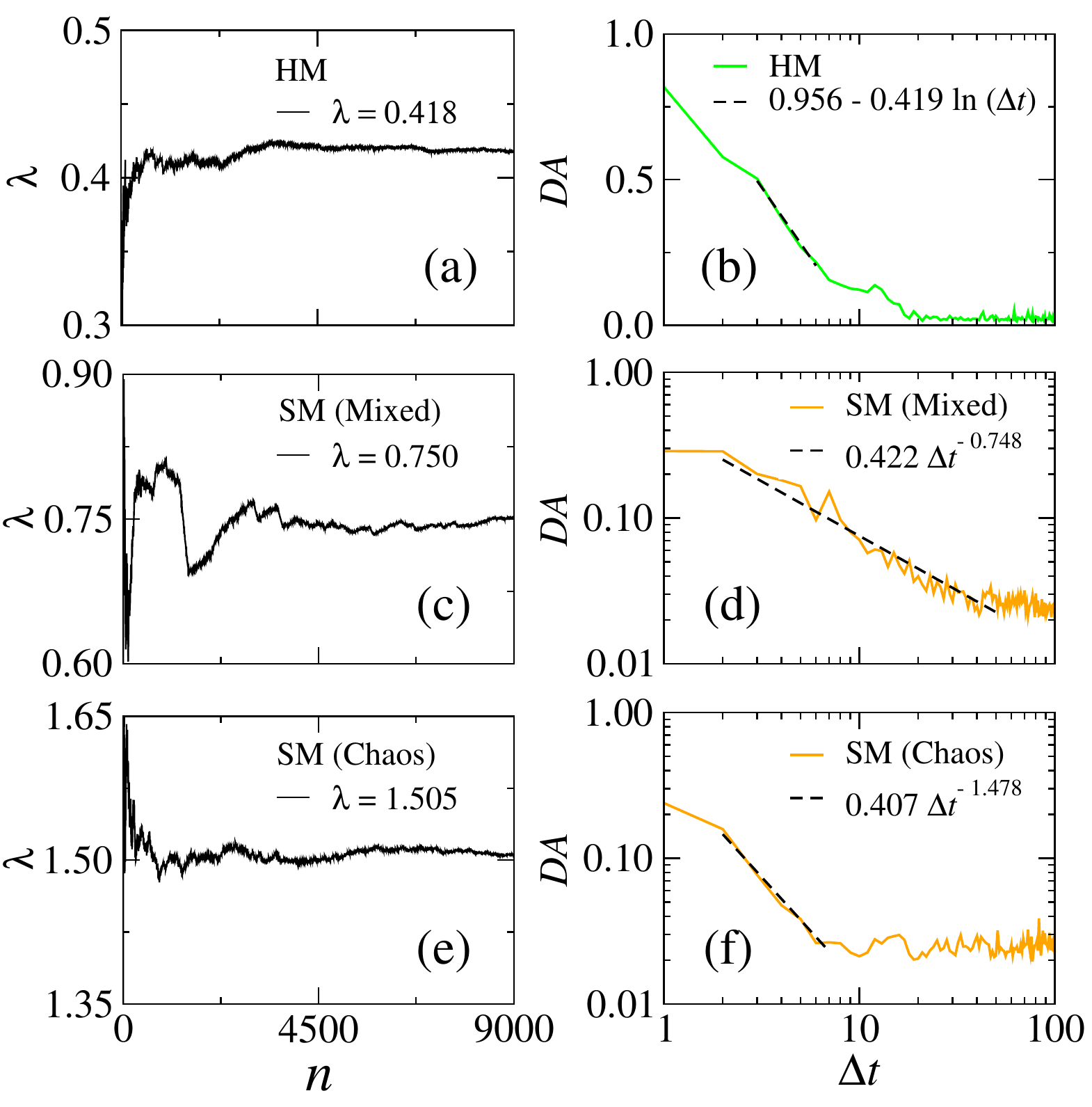}
\caption{(a) LE and (b) the corresponding $DA(\Delta t)$ (green curve) in semi-log scale 
for the HM (\ref{Hmap}) with parameters $r=1.4$ and $b=0.3$. In (c) and (e), the LE for the 
SM (\ref{stand-map}) with mixed and chaotic phase space are displayed, respectively. The figures 
(d) and (f) show the corresponding $DA(\Delta t)$ (orange curve) in the log-log scale. The 
dashed black lines in (b), (d) and (f) are the associated adjustments.}
\label{fig7}
\end{center}
\end{figure}

{\it H\'enon map} (HM).
The H\'enon map is a generalization of the quadratic map for two dimensional systems, and it 
is given by \cite{henon76}
\begin{equation}
\label{Hmap} 
\begin{array}{ll}
x_{n+1} = r-x_{n}^{2} + b y_n,\\
y_{n+1} = x_n.
\end{array}
\end{equation}
The additional parameter $b$ determines the dissipation of the systems. The HM is dissipative 
for $|b|<1$. This two-dimensional map has two LEs. It is well known that the HM has a chaotic 
attractor for $r=1.4$ and $b=0.3$ \cite{henon76} with positive LE $\lambda=0.418$, as shown 
in Fig. \ref{fig7}(a). Since we are considering two dimensions, the quantities for $DA$
are obtained from {$|X_i-X_{j}| =\sqrt{(x_i-x_{j})^2+(y_i-y_{j})^2}$}. 
Results are displayed in Fig.~\ref{fig7}(b) and show the corresponding behavior of $DA$. 
{When we refer to the adjusted curve we use the notation $DA(\Delta t)$.} As seen in this 
figure, the decay of $DA$ is well adjusted by the function $DA(\Delta t)\sim0.956-0.419\hspace{0.1cm}
\mbox{ln}(\Delta t)$. Thus the time decay coefficient $0.419$ is close to the LE 
$\lambda=0.418$.

{\it Standard map} (SM). Here we consider the two-dimensional conservative 
SM given by \cite{sm-chirikov}
\renewcommand{\arraystretch}{1.1}
\begin{equation}
\label{stand-map}
\begin{array}{ll}
p_{n+1} = p_{n} + K \sin(2\pi \hspace{0.02cm} x_{n}) &  \hspace{0.4cm} [\mathrm{mod} 
\hspace{0.3cm} 1], \\
x_{n+1} = x_{n} + p_{n+1} & \hspace{0.4cm} [\mathrm{mod} \hspace{0.3cm} 1], \\
\end{array}
\end{equation}
\noindent where $x_n$ is the position at the iteration $n=0,1,2,\ldots$, and $p_n$ 
its conjugated momentum. $K$ is the nonlinear positive parameter, and its
value determines the topology of the phase space. Some values of $K$ lead to 
a mixed phase space compound by Kolmogorov-Arnold-Moser (KAM) tori and a 
stochastic region  \cite{licht}. The chaotic trajectory can be trapped for long, but 
finite times  close to the tori, given origin to the {\it stickiness effect} 
\cite{mo5527411985}. This scenario can be obtained using $K=0.57$, for which $\lambda=
0.750$, as shown in Fig. \ref{fig7}(c). A totally chaotic phase space takes place if 
$K=1.43$, with $\lambda=1.505$ as displayed in Fig. \ref{fig7}(e). In both cases the 
appropriate adjustment was $DA(\Delta t)\propto \alpha \Delta t^{-\beta}$. Also here the 
quantities for $DA$ are obtained from $|X_i-X_j| =\sqrt{(x_i-x_j)^2+(p_i-p_j)^2}$. For 
the mixed case we have $\alpha=0.422$ and $\beta=0.748$, which is very close to the 
associated LE [see Fig.~\ref{fig7}(d)], and for the chaotic case we found $\alpha=0.407$ 
and $\beta=1.478$ [Fig.~\ref{fig7}(f)], which is also close to the respective LE 
$\lambda=1.505$. 

{\it Coupled Standard Maps} (CSMs). In this model we consider a coupling between 
identical conservative SMs. For our numerical investigation we used the $2$-dimensional 
SM \cite{edu07,silva15}:  
\renewcommand{\arraystretch}{1.3}
\begin{equation}
  \label{mp-acop1}
  \mathbf{M_i}\left(
  \begin{array}{c}
    p_i \\
    x_i \\
  \end{array}
  \right) = \left(
  \begin{array}{llll}
    p_i + K_i \sin(2\pi x_i) & \hspace{0.1cm} [\mathrm{mod} \hspace{0.2cm} 1] \\
    x_i + p_i + K_i \sin(2\pi x_i) & \hspace{0.1cm} [\mathrm{mod} \hspace{0.2cm} 1] \\
  \end{array}
  \right),
\end{equation}
\noindent and for the coupling 
\begin{equation}
  \label{mp-acop2}
  \mathbf{T_i} \left(
  \begin{array}{c}
    p_i \\
    x_i \\
  \end{array}
  \right) = \left(
  \begin{array}{llll}
    p_i + \sum_{j=1}^{N} \xi_{i,j} \hspace{0.05cm} \sin[2\pi (x_i - x_j)] \\
    x_i \\
  \end{array}
  \right),
\end{equation}
\noindent with $\xi_{i,j} = \xi_{j,i} = \frac{\xi}{\sqrt{N-1}}$ (all-to-all coupling). 
This constitutes a $2N$-dimensional Hamiltonian system and the total map
is a composition of $T$ and $M$. We considered two cases: (i) $N=2$ 
and (ii) $N=4$, for which $DA$ is obtained from 
$$|X_i-X_j| =\sqrt{\sum_k^N\left[
\left(x^{(k)}_i-x^{(k)}_j\right)^2+\left(p^{(k)}_i-p^{(k)}_j\right)^2\right]}.$$

\begin{figure}[!t]
\begin{center} 
  \includegraphics[width=8.7cm,height=8.2cm]{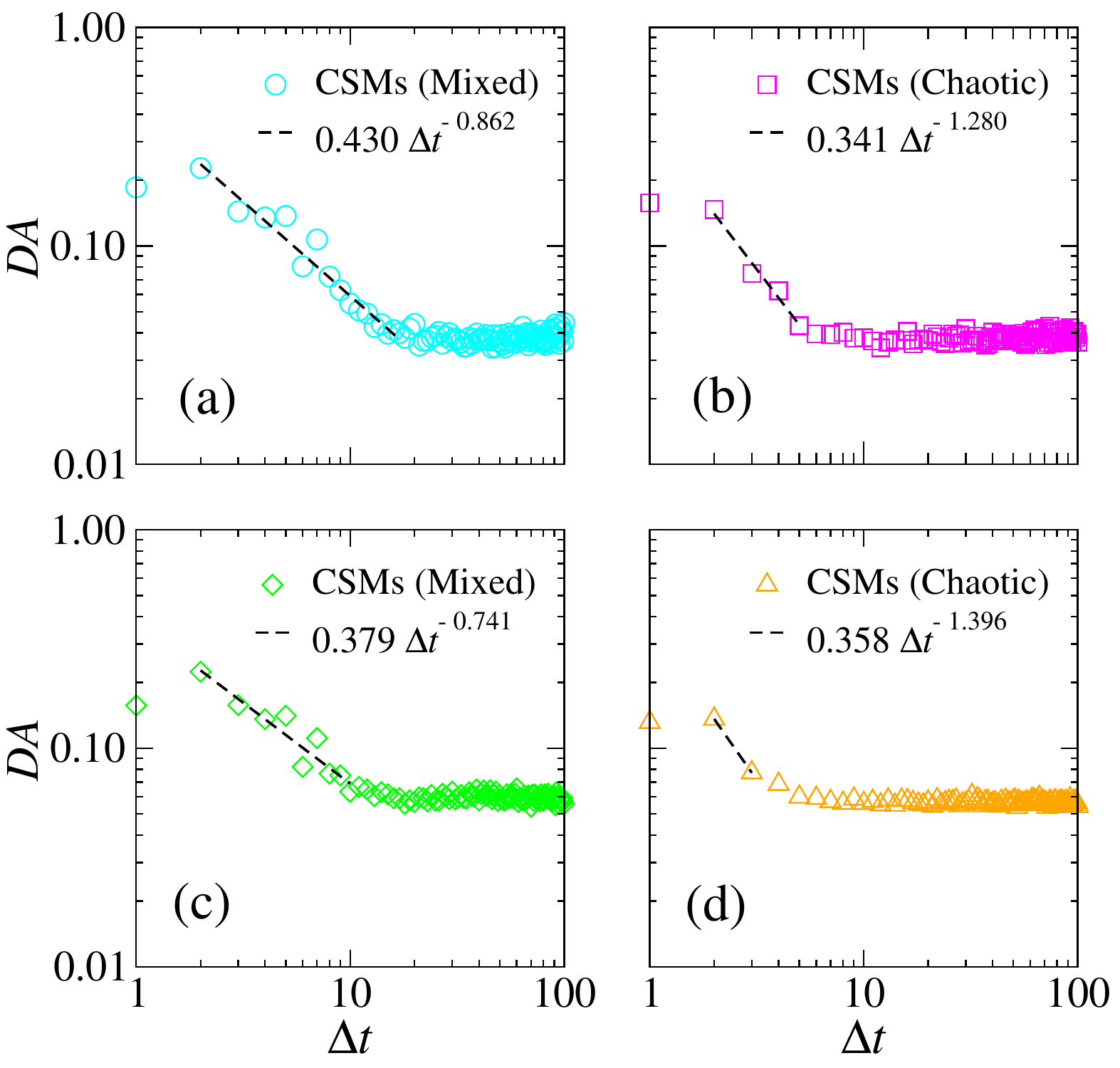}
  \caption{$DA(\Delta t)$ for CSMs (\ref{mp-acop1})-(\ref{mp-acop2}) with $N=2$ (a) 
  for the mixed (circles in cyan) and (b) for the chaotic (squares in magenta) cases. With 
  $N=4$, the mixed (green diamonds) and the chaotic (orange triangles) cases are displayed in 
  figures (c) and (d), respectively. The dashed-black lines are the associated adjustment. 
  Results are summarized in Table I.}
\label{fig8}
\end{center}
\end{figure}

{\it Case (i).} Here we have two positive LEs ($\lambda_1=0.860$ and 
$\lambda_2=0.727$) for the mixed dynamics obtained using $K_1=0.57$ and
$K_2=0.59$. Calculating $DA(\Delta t)$ for this case, we obtain $\beta = 0.862$ 
[see Fig. \ref{fig8}(a)]. Considering the chaotic case with $K_1=1.43$ 
and $K_2=1.45$, we have LEs $\lambda_1=1.563$ and $\lambda_2=1.469$ and
the parameter that fits the decay of $DA(\Delta t)$ is
$\beta=1.280$, as seen in Fig. \ref{fig8}(b). In either case the function
used to the adjustment was $DA(\Delta t)\sim \alpha \Delta t^{-\beta}$.
\begin{table}[hb]
\caption{Table presents the models and corresponding positive LEs $\lambda_i (i=1,2,3,4)$ 
with the adjustment  curves and the associated decay exponents $\beta$.}
\begin{tabular}{|c|c|c|c|c|c|c|}\hline
Model           & $\lambda_1$ & $\lambda_2$ & $\lambda_3$ & $\lambda_4$ & $\beta$ & $DA(\Delta t)$ \\ \hline\hline
HM              & 0.418 &   -       &    - & - &  0.419  &   $-\beta\, \mbox{ln}(\Delta t)$ \\  \hline
SM (mixed)      & 0.750 &   -       &    - & - & 0.748       &   $0.422\,\Delta t^{-\beta}$ \\  
SM (chaotic)    & 1.505 &   -        &     -& - &1.478       &   $0.407\,\Delta t^{-\beta}$ \\  \hline
2-CSM (mixed)   & 0.860 & 0.727 &   -      &  -& 0.862 &$0.430\,\Delta t^{-\beta}$ \\  
2-CSM (chaotic) & 1.563 & 1.469 &   -      &  -& 1.280&$0.341\,\Delta t^{-\beta}$ \\  \hline
4-CSM (mixed)   & 0.922 & 0.853 & 0.776      &  0.624& 0.741 &$0.379\,\Delta t^{-\beta}$ \\  
4-CSM (chaotic) & 1.608 & 1.558 & 1.500      &  1.437 & 1.396&$0.358\,\Delta t^{-\beta}$ \\  \hline
\end{tabular}
\end{table}

{\it Case (ii)}. Here we have four positive LEs ($\lambda_1=0.922,\lambda_2=0.853,\lambda_3=0.776$ 
and $\lambda_4=0.624$) for the mixed dynamics obtained with $K_1=0.57$, $K_2=0.58$, $K_3=0.59$ 
and $K_4=0.60$. In this case, again we found for the adjustment the function $DA(\Delta t)=
\alpha\,\Delta t^{-\beta}$, with $\beta=0.741$, as can be seen in the Fig. \ref{fig8}(c). For the chaotic case, 
obtained with $K_1=1.42$,  $K_2=1.43$, $K_3=1.44$ and $K_4=1.45$, we have LEs 
$\lambda_1=1.608, \lambda_2=1.558, \lambda_3=1.500$ and $\lambda_4=1.437$. 
Results are shown in Fig. \ref{fig8}(d) and we obtain $\beta=1.396$.
 
Results for higher-dimensional systems are summarized in Table I. The decay of $DA$ in 
higher-dimensional systems with more than one positive LE does not give precise information 
about the LEs. {In such cases $\beta$ is close to the smallest LE, a feature also observed
for the autocorrelation function \cite{collet04}}.


\section{Conclusions}
\label{conclusions}

The present work analyzes numerically the relation between the decay of the distance
autocorrelation and LEs. 
Results are shown for the dissipative quadratic and H\'enon map and for the conservative
standard maps and coupled standard maps. For all conservative cases we found a decay 
for $DA$ proportional to $\Delta t^{-\beta}$, where $\beta$ is very close to the LE
$\lambda$ in case the system has just one positive LE. For hyperchaotic systems, 
the parameters $\beta$ tends to follow the smallest LE {(with one exception)}. 
For the 1D dissipative system  (quadratic map) the observed decay obeys $e^{-\beta \Delta 
t}$ with $\beta\sim\lambda$. For the 2D dissipative system (H\'enon map), the decay follows 
$-\beta\, \mbox{ln}(\Delta t)$, and $\beta\sim\lambda$. Thus, for systems with one 
positive LE the decay of the distance autocorrelation is nicely related to the 
Lyapunov exponent.   

Further investigations will study in more details the decay of $DA$ in hyperchaotic  and in 
time continuous systems. Moreover, it should be possible to extract from $DA$ more 
general aspects of the dynamics, since $DA$ certainly contains information which 
go beyond the {linear stability analysis} of the exponential divergence of trajectories, 
quantified  by the LE.

\section*{acknowledgments}

CFOM thanks FAPEAM, RMS thanks CAPES and MWB thanks CNPq for financial support. The authors 
also acknowledge computational support from Professor Carlos M.~de Carvalho at LFTC-DFis-UFPR.
\section*{References}

\end{document}